\documentclass[twocolumn,tighten,times,twocolappendix,numberedappendix]{aastex631}
\usepackage{CJK}
\usepackage{natbib}

\usepackage{xcolor}

\received{\today}

\submitjournal{ApJL}

\begin{document}
\title{\large XMM-Newton Detection of X-ray Emission from the Metal-Polluted White Dwarf G\,29-38}

\begin{CJK*}{UTF8}{gbsn}

\author[0000-0002-7563-4513]{S.~Estrada-Dorado}
\affil{Instituto de Radioastronom\'{i}a y Astrof\'{i}sica, UNAM, Ant. carretera a P\'{a}tzcuaro 8701, Ex-Hda. San Jos\'{e} de la Huerta, Morelia 58089, Mich., Mexico}

\author[0000-0002-7759-106X]{M.~A.\ Guerrero}
\affiliation{Instituto de Astrof\'{i}sica de Andaluc\'{i}a, IAA-CSIC, Glorieta de la Astronom\'{i}a s/n, Granada E-18008, Spain}

\author[0000-0002-5406-0813]{J.~A.~Toal\'{a}\,(杜宇君)}
\altaffiliation{Visiting astronomer at the Instituto de Astrof\'{i}sica de Andaluc\'{i}a (IAA-CSIC, Spain) as part of the Centro de Excelencia Severo Ochoa Visiting-Incoming programme.}
\affil{Instituto de Radioastronom\'{i}a y Astrof\'{i}sica, UNAM, Ant. carretera a P\'{a}tzcuaro 8701, Ex-Hda. San Jos\'{e} de la Huerta, Morelia 58089, Mich., Mexico}

\author[0000-0003-3667-574X]{Y.-H.\,Chu\,(朱有花)}
\affil{Institute of Astronomy and Astrophysics, Academia Sinica, No. 1, Section 4, Roosevelt Road, Taipei 10617, Taiwan}

\author[0000-0003-3588-5235]{V.\,Lora}
\affil{Instituto de Radioastronom\'{i}a y Astrof\'{i}sica, UNAM, Ant. carretera a P\'{a}tzcuaro 8701, Ex-Hda. San Jos\'{e} de la Huerta, Morelia 58089, Mich., Mexico}
\affil{Instituto de Ciencias Nucleares, UNAM, Apartado postal 73-543, 04510 Ciudad de M\'{e}xico, Mexico}

\author[0000-0001-5559-7850]{C.~Rodr\'{i}guez-L\'{o}pez}
\affiliation{Instituto de Astrof\'{i}sica de Andaluc\'{i}a, IAA-CSIC, Glorieta de la Astronom\'{i}a s/n, Granada E-18008, Spain}



\begin{abstract}
A recent analysis of Chandra X-ray data of the metal-polluted white dwarf (WD) G\,29-38 has revealed X-ray emission that can be attributed to the accretion of debris from a planetary body. In the light of this detection we revisit here archival XMM-Newton observations of G\,29-38 from which only an upper limit was derived in the past due to the presence of a relatively bright nearby X-ray source. An analysis of these data in multiple energy bands allows disentangling the X-ray emission at the location of G\,29-38 from that of the nearby source. The similar spectral properties of the source in the XMM-Newton and Chandra observations and their spatial shift, consistent with the proper motion of G\,29-38 between these observations, strengthen the origin of that X-ray emission from G\,29-38.  
The X-ray luminosities from both observations are consistent within 1-$\sigma$ uncertainties, so too are the best-fit plasma temperatures. Although the count number is small, there is tantalizing evidence for line emission in the 0.7--0.8 keV energy band from an optically-thin hot plasma. The most likely candidate for this line emission would be the Fe complex at 16 \AA.
\end{abstract}

\keywords{\href{https://astrothesaurus.org/uat/1799}{White dwarf stars(1799)};
\href{http://astrothesaurus.org/uat/2050}{Stellar accretion(1578)};
\href{http://astrothesaurus.org/uat/2050}{X-ray stars(1823)};
\href{http://astrothesaurus.org/uat/2050}{Low mass stars(2050)}
\vspace{4pt}
\newline
}


\section{Introduction}
\end{CJK*}
\label{sec:intro}

The late evolution of white dwarfs (WDs), the stellar end products of low- and intermediate-mass stars, can be described as a long-lasting cooling process as their thermal energy is radiated away \citep[e.g.,][]{Renedo_etal2010}. By the time ($\simeq$20 Myr) a WD's temperature falls below $\simeq$25,000 K,
metals are no longer radiatively supported in the high gravity atmosphere and sink below the surface \citep{Chayer1995a,Chayer1995b,Koester2014}, with diffusion time-scales from days to a few years for H-rich atmospheres \citep{KKI2020}.
Therefore the presence of metallic absorption lines in the spectra of a cool degenerate WDs implies that they have accreted metal-rich material after their formation, most likely from circumstellar debris disks evidenced by infrared excess \citep{Jura2003,Jura2008,Farihi_etal2010,HGK2018}. This way metal-polluted WDs provide unique means to investigate the late fate of planetary systems and to determine their bulk abundances \citep[e.g.,][]{Gansicke_etal2012}.

\citet{Cunningham2022} (hereafter C2022) has recently presented Chandra observations of G\,29-38 (a.k.a.\ WD\,2326$+$049), which resulted in the detection of 5 photons, implying an X-ray luminosity $\approx$10$^{26}$~erg~s$^{-1}$ \citep[at the distance of 17.53$\pm$0.01~pc,][]{Gaia2020}.  
Since G\,29-38 is a single cool degenerate WD, the origin of these X-rays can be excluded to be (i) the photospheric emission of a hot WD, (ii) the coronal emission from a late-type companion, or (iii) the emission from a close binary companion accretion disk \citep{Chu2021}.

G\,29-38 was identified as the first metal-polluted WD \citep{KPS1997} of the spectral subtype DAZ \citep{Zuckerman_etal2003}, i.e., its atmosphere includes absorption lines both of H and a number of metals \citep[mostly Ca, Mg, and Fe,][]{Xu_etal2014}.  
It is also the first metal-polluted WD where infrared excess was identified \citep{ZB1987}.  
This was interpreted later as an orbiting dust disk attributed to the disruption of an asteroid or a minor planet \citep{Jura2003,Reach_etal2005}.

The accretion of material from this disk onto G\,29-38 would produce the X-ray emission \citep{KL1982} detected by C2022. Very interestingly G\,29-38 was observed in 2005 by XMM-Newton, which brings up the tantalizing possibility to investigate its X-ray variability. 
\citet[][]{Jura2009} (hereafter J2009) originally analyzed that data set and reported an upper limit for its X-ray flux about two times smaller than that reported by C2022. 
J2009 remarked, however, that their upper limit was conservatively adopted given the presence of a nearby brighter background X-ray source.  
A more recent analysis of the same XMM-Newton data limited  to its MOS cameras raised the upper limit of J2009 by a factor of three \citep[][hereafter F2018]{Farihi_etal2018}, although it was attributed to contamination of the nearby background source.

The differing values of the X-ray upper limits derived from the XMM-Newton data of G\,29-38 reported by J2009 and F2018 and the Chandra detection reported by C2022 certainly recommend the need for a re-analysis of the XMM-Newton data set. In this letter, guided by the recent Chandra detection of X-ray emission from G\,29-38, we revisit its XMM-Newton observations to evaluate whether it was indeed detected or its X-ray flux has varied. 
\\

\section{Observations and data preparation}
\label{sec:obs}

We analyze the XMM-Newton observations of G\,29-38 obtained on 2005 November 28 (Obs.~ID 0302820101; PI: M.~Muno) with total exposure time of 24.8~ks. 
The observations were processed using the Science Analysis Software \citep[SAS, version 18.0.0;][]{Gabriel2004}.
Event files of the European Photon Imaging Cameras (EPIC) pn and MOS cameras were created using the {\it epproc} and {\it emproc} SAS tasks, respectively. 
Time periods with background count rates in the 10--12 keV energy range above 0.15 counts s$^{-1}$ for the MOS cameras and 0.4 counts s$^{-1}$ for the EPIC pn were excised. 
The net exposure time for the EPIC-pn, which will be otherwise used in the following, is 17.2 ks.

To facilitate the investigation of the spatial distribution of X-rays in the region around G\,29-38, very particularly to separate its emission from that of the nearby background source, we used the Extended Source Analysis Software (ESAS) package \citep{Snowden2004,Snowden2008,Kuntz2008} to create exposure-corrected, background-subtracted images in different energy ranges. 
We note that the ESAS tasks have quite restrictive event selection criteria, resulting in a lower EPIC-pn net exposure time of 16.8 ks, but they leverage the presence of extended and point-like sources. 
EPIC-pn images were then created in the 0.3-1.0 keV (soft) and 1.0-2.0~keV (hard) energy bands. 
After the spectral analysis (see Section~\ref{sec:spec_analysis}), evidence was found for line emission between 0.7 and 0.8 keV from G\,29-38, which is otherwise not especially bright in the nearby background source. Thus an additional EPIC-pn image in this narrow energy range was produced to accentuate the separation between G\,29-38 and the nearby X-ray source. The three images were adaptively smoothed using the ESAS task {\it adapt} requesting a minimum of 5 counts.

\begin{figure}
\begin{center}
\includegraphics[angle=0,width=0.93\linewidth]{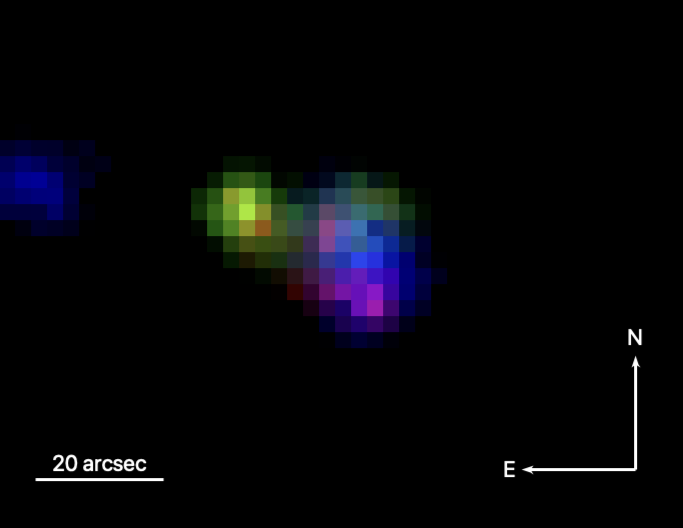}

\vspace*{0.1cm}
\includegraphics[angle=0,width=0.93\linewidth]{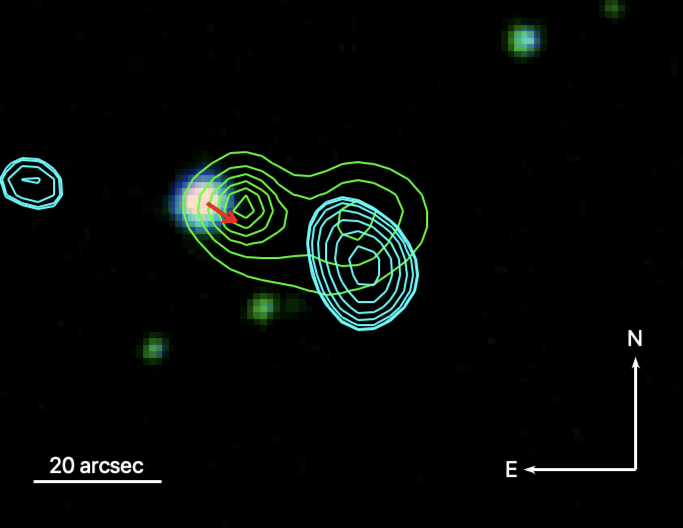} 

\vspace*{0.1cm}
\includegraphics[angle=0,width=0.93\linewidth]{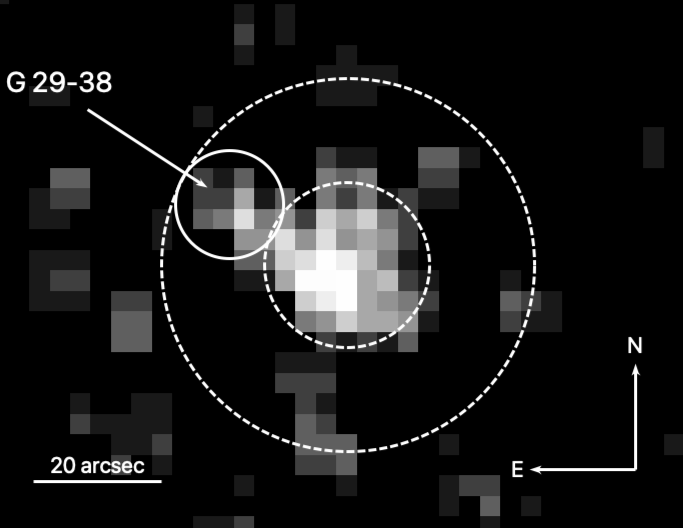}
\caption{X-ray and optical/IR views of G\,29-38. {\it top}: ESAS EPIC X-ray composite-picture in the 0.3--1.0 keV (red), 0.7--0.8 keV (green), and 1.0--2.0 keV (blue) energy bands. {\it middle}: DSS Red, Blue and IR color-composite picture (1993 August 22) overimposed by X-ray contours in the 0.7--0.8 keV (green) and 1.0--2.0 keV (cyan) bands. The red arrow shows the location of G\,29-38 at the epoch of the XMM-Newton observation (2005.91). {\it bottom}: EPIC-pn event image with a solid circle showing the aperture used to extract the spectrum of G\,29-38, and inner and outer dashed circles representing the circular aperture used to extract the spectra of the nearby source and the annular aperture used to extract a background spectrum suitable for G\,9-38. 
}
\label{fig:region}
\end{center}
\end{figure}

\begin{figure*}
\begin{center}
\includegraphics[angle=0,width=\linewidth]{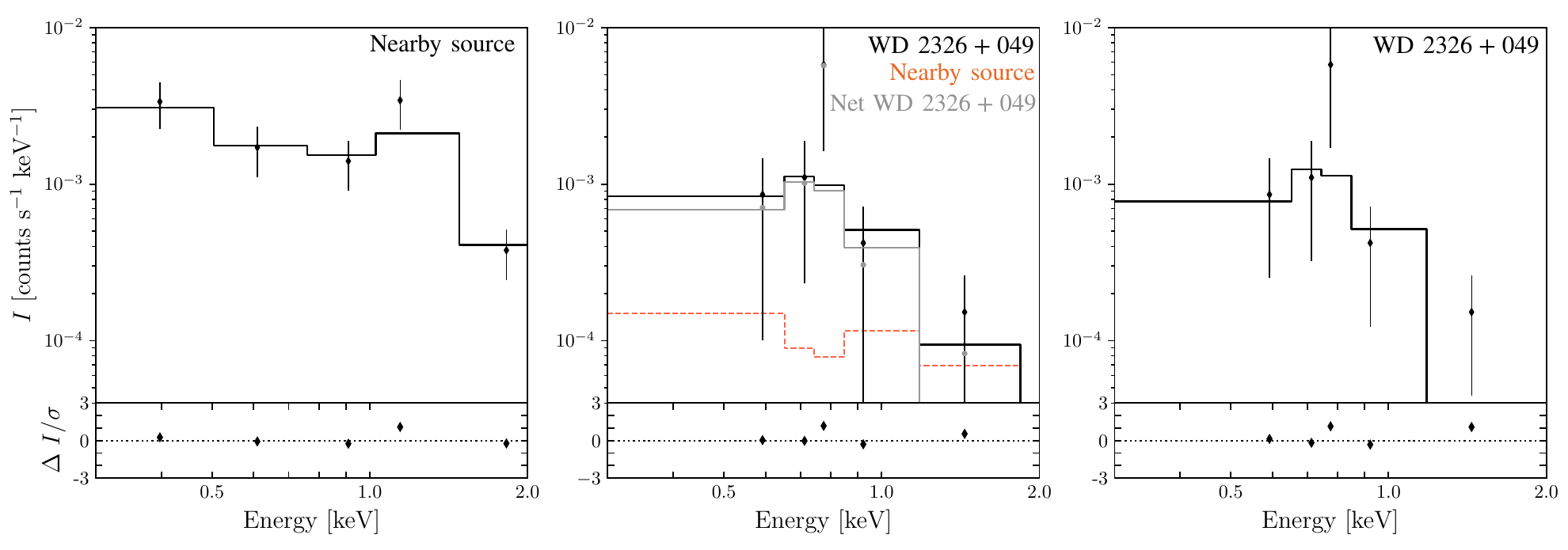}
\caption{{\it left}: XMM-Newton EPIC-pn background-subtracted spectra (dots) of the source nearby to G\,29-38. The black solid histogram represents the best-fit to the data and the lower panel the residuals of the fit. {\it center}: XMM-Newton EPIC-pn background-subtracted spectra (dots) of G\,29-38 using distant background regions.  
The red-dashed histogram describes the contribution of the nearby source, whereas the grey histogram corresponds to the X-ray model of G\,29-38 described by C2022.  
{\it right}: XMM-Newton EPIC-pn background-subtracted spectra (dots) of G\,29-38 using a background region around the nearby source. The black histogram is the optically-thin isothermal ($kT$=0.49 keV) {\it vapec} plasma X-ray emission model of G\,29-38 with chemical abundances of the WD photosphere described by C2022. All spectra are binned to 2 counts per spectral bin.}
\label{fig:spectrum}
\end{center}
\end{figure*}

\section{Data Analysis}

\subsection{Source Identification}

In their analysis of this same XMM-Newton dataset, J2009 reported the presence of a background X-ray source only $\approx15^{\prime\prime}$ from the expected location of G\,29-38.  
This background source, being brighter than any X-ray emission from the location of G\,29-38, certainly complicates its identification.  
The recent Chandra detection of X-ray emission from G\,29-38 reported by C2022 clarifies its relative position with respect to background X-ray sources.  
The use of the ESAS task to produce a color-picture of the region around G\,29-38 in the energy bands 0.3--1.0, 0.7--0.8, and 1.0--2.0 keV indeed confirms the presence of two distinct X-ray sources, one fainter and ``yellow'' and another brighter and ``purple'' $\approx20^{\prime\prime}$ to its Southwest (Figure~\ref{fig:region}-{\it top}).

To assess whether any of these X-ray sources is associated to G\,29-38, we retrieved optical and near-IR images in the POSS2/UKSTU Blue, Red and IR bands from the Digitized Sky Survey (DSS)\footnote{\url{https://archive.stsci.edu/cgi-bin/dss_form/}} and present it in Figure~\ref{fig:region}-{\it middle} overlaid by X-ray contours in the 0.7--0.8 and 1.0--2.0 keV bands that emphasize the emission from each of these sources. 
The J2000.0 coordinates of the peak emission of these sources are 23:28:47.39, +05:14:55.0 and 23:28:46.14, +05:14:46.1, respectively. 
The ``yellow'' source is actually located 2.5$''$ from the expected position of G\,29-38, once its large proper motion \citep[$\delta_\alpha = -398\farcs246\pm0\farcs032$~yr$^{-1}$, 
$\delta_\delta  = -266\farcs744\pm0\farcs020$~yr$^{-1}$,][]{Gaia2020} 
illustrated in Figure~\ref{fig:region}-{\it middle} by a red arrow is considered to compute the shift of its position from the DSS image (1993) to locate its position at the epoch of the XMM-Newton observation (2005.91).  
This offset is within the XMM-Newton spatial resolution.

\subsection{Spectra Extraction}

The XMM-Newton EPIC-pn point spread function has a half energy width (HEW) of 16\farcs6, resulting in non-negligible contribution of the X-ray emission from the source nearby to the location of G\,29-38\footnote{
As illustrated in the middle panel of Figure~\ref{fig:region}, the projected location of G\,29-38 moves closer to that of the background X-ray source as time proceeds.  
Indeed their distance has reduced to $\simeq$8\farcs4 at the time of the Chandra observation in 2020.  
The earlier observation by XMM-Newton, when the sources separation was $\approx21^{\prime\prime}$, is thus a fortunate fact that eases the separation of their respective emissions. 
}.  
To reduce and to assess this contamination, we used a circular aperture 8\farcs3 in radius to extract the spectrum of G\,29-38 (Fig.~\ref{fig:region}-{\it bottom}), together with a suitable background region with an area $\simeq$40 times larger than the source aperture consisting of several nearby circular apertures free from sources. 
Since the source aperture encompasses the EPIC-pn HEW, the encircled energy fraction in the background-subtracted spectrum shown in Fig.~\ref{fig:spectrum}-{\it center} is 0.5.  
We then proceeded to extract the spectrum of the nearby source and fitted its background-subtracted spectrum (Fig.~\ref{fig:spectrum}-{\it left}) with a suitable model.  
It is important to note the spectral differences between the nearby source and the WD suggested by the color-composite picture in Fig.~\ref{fig:region}-{\it top} and confirmed by their spectra presented in the left and center panels of Fig.~\ref{fig:spectrum}. 
The EPIC-pn encircled energy fraction was used to compute the contribution of this source to the aperture used for G\,29-38 and the spectrum of the nearby source adequately scaled subtracted from the spectrum of G\,29-38 (Fig.~\ref{fig:spectrum}-{\it center}).  
The net EPIC-pn count number from G\,29-38 is 9$\pm$3 cnts that, after accounting for the encircled energy fraction of 0.5, corresponds to a count rate of 1.0$\pm$0.4 cnts~ks$^{-1}$.

Alternatively, we have selected a background region for G\,29-38 from an annular region around the nearby source in the radius range of the source region (Figure~\ref{fig:region}-{\it bottom}).  
This background spectrum thus accounts for the expected contribution of this source to G\,29-38.  
The comparison between this spectrum and the net spectrum of G\,29-38 in the right and center panels of Fig.~\ref{fig:spectrum} shows noticeable agreement.   
Indeed, the net EPIC-pn count number and PSF-corrected count rate in this spectrum, 10$\pm$3 cnts and 1.2$\pm$0.4 cnts~ks$^{-1}$, respectively, are consistent within the uncertainties with those derived in the paragraph above.
For comparison, the Chandra ACIS-S count rate reported by C2022 in the 0.5--2.0 keV band is 0.047$^{+0.023}_{-0.020}$ cnts~ks$^{-1}$, whereas the XMM-Newton EPIC-pn and EPIC-MOS count rate 3$\sigma$ upper limits in the 0.3--2.0 keV band reported by J2009 are $<$0.9 cnts~ks$^{-1}$ and $<$0.32 cnts~ks$^{-1}$, 
respectively.
Meanwhile F2018 reported an XMM-Newton EPIC-MOS count rate 3$\sigma$ upper limit in the 0.3--2.0 keV band of $<$0.8 cnts~ks$^{-1}$.

\subsection{Spectral Analysis}
\label{sec:spec_analysis}

The total count number derived from the EPIC-pn background-subtracted spectra of G\,29-38 in the center and right panels of Fig.~\ref{fig:spectrum}-right is obviously too small to allow a detailed spectral modeling.  
Instead we compare these spectra using the XSPEC package \citep[version 12.10.1;][]{Arnaud1996} with the thin-plasma emission model described by C2022 consisting of an optically-thin plasma emission model with plasma temperature of 0.49 keV and chemical abundances of the WD photosphere \citep[as described in Table~3 of][]{Xu_etal2014} absorbed by a hydrogen column density $N_\mathrm{H}$ of 5.4$\times10^{18}$ cm$^{-2}$.  
Calibration matrices were obtained using the standard \emph{rmfgen} and \emph{arfgen} SAS tasks. 
The \emph{tbabs} absorption component \citep{Wilms2000} was adopted together with the variable abundances \emph{vapec} model. 
We note that the adopted value of $N_\mathrm{H}$ is about 100 times smaller than the value provided by the NASA's HEASARC NH column density tool \citep[HI4PI2016, ][]{Kalberla2005, Dickey1990}\footnote{\url{https://heasarc.gsfc.nasa.gov/cgi-bin/Tools/w3nh/w3nh.pl}}, but it most likely represents the small absortion towards this nearby WD. 
This model has an X-ray flux and luminosity of $F_\mathrm{X}=(2.2\pm1.1)\times10^{-15}$ erg~cm$^{-2}$~s$^{-1}$ and $L_\mathrm{X}=(8.3\pm4.1)\times10^{25}$~erg~s$^{-1}$, respectively, which are consistent with those estimated by C2022 for the 0.3 to 7.0 keV band.
The model provides a reasonable description of the XMM-Newton data\footnote{We note that, although the source aperture used to extract the spectrum of G\,29-38 only includes 50\% of its emission, the calibration matrices correct the emission from the incomplete coverage of the source PSF.}, but it seems to underestimate the observed X-ray emission at $\approx$0.8 keV. 
To illustrate further the differences between our results and those presented by C2022, we list in Table~\ref{tab:parameters} the model parameters used for the spectral analysis of the XMM-Newton and Chandra data, where we emphasize that our X-ray temperature was fixed to that derived by C2022.

\begin{table}
\begin{center}
\caption{Observed properties of G\,29-38 obtained from the XMM-Newton (this work) and Chandra (from C2022) observations. The X-ray flux ($F_\mathrm{X}$) and luminosity ($L_\mathrm{X}$) were computed for the 0.3--2.0~keV energy range.}
\begin{tabular}{lcc}
\hline
       & XMM-Newton & Chandra \\
\hline
$T_\mathrm{X}$ [keV] & 0.49  & 0.49$^{+0.17}_{-0.21}$  \\
$T_\mathrm{X}$ [K] & 5.7$\times10^{6}$ & (5.7$^{+2.0}_{-2.5}$)$\times10^{6}$\\
$F_\mathrm{X}$ [erg~cm$^{-2}$~s$^{-1}$] & (2.2$\pm$1.1)$\times$10$^{-15}$ & (2.0$^{+1.6}_{-0.5}$)$\times$10$^{-15}$\\
$L_\mathrm{X}$ [erg~s$^{-1}$] & (8.3$\pm$4.1)$\times$10$^{25}$ & (7.2$^{+5.7}_{-1.8}$)$\times$10$^{25}$  \\
\hline
\end{tabular}
\label{tab:parameters}
\end{center}
\end{table}

\begin{figure}
\begin{center}
\includegraphics[bb=45 210 570 580,angle=0,width=\linewidth]{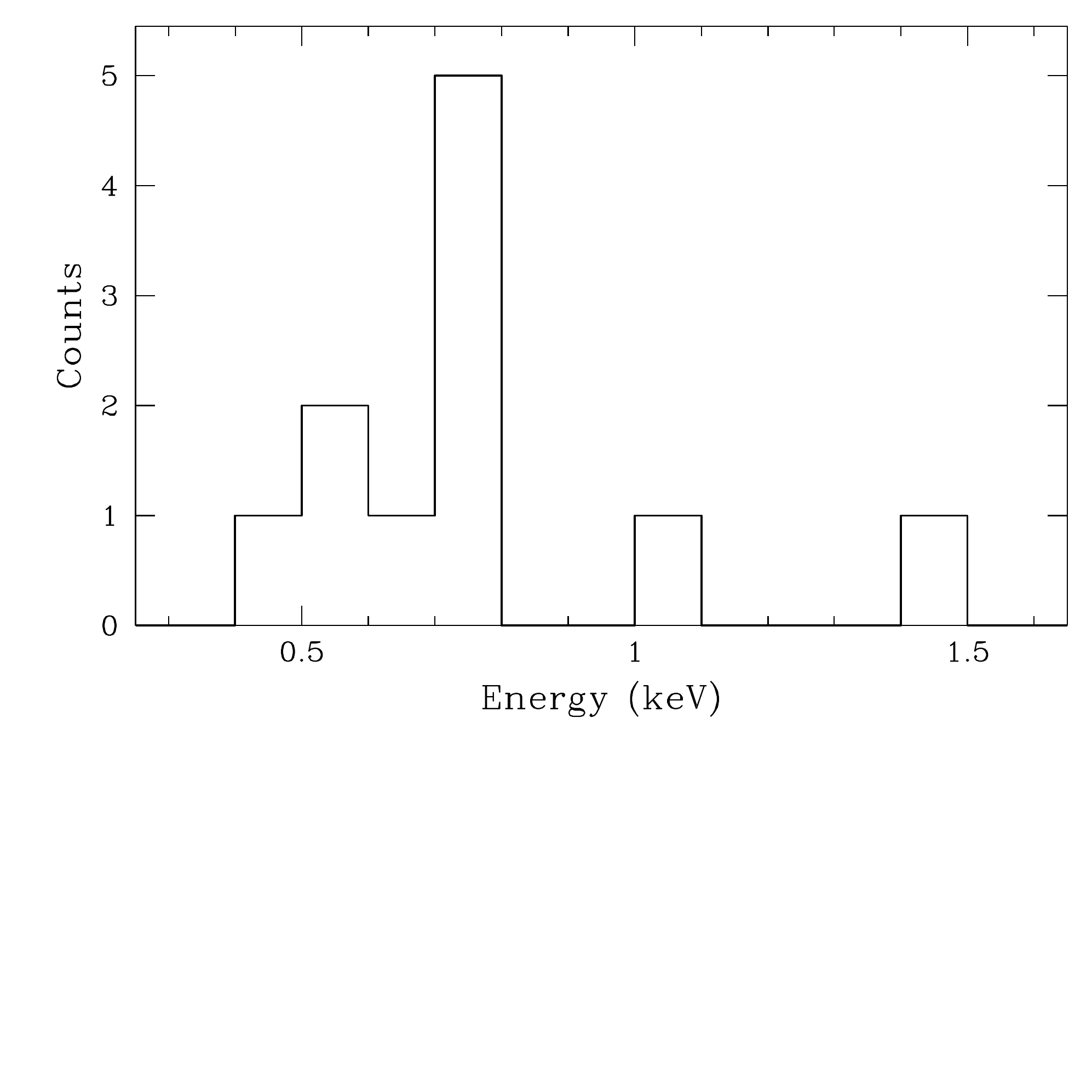}
\caption{XMM-Newton EPIC-pn background-subtracted spectrum of G\,29-38 with a bin width of 0.1 keV.} 
\label{fig:fe_line}
\end{center}
\end{figure}

To further investigate this apparent excess, we present in Figure~\ref{fig:fe_line} the unbinned EPIC-pn background-subtracted spectrum of G\,29-38.  
This is basically dominated by the emission in the 0.7--0.8 keV range, with 6 out of the 10 counts in this energy range.  
The emission from G\,29-38 in the 0.7-0.8 keV band even outshines that of the nearby source (Fig.~\ref{fig:region}-top). 
This spectral behavior is also the case for the Chandra spectrum, with 4 out of its 5 counts in the 0.7 to 1.0 keV energy range (C2022).  
The combined Chandra and XMM-Newton detection of 10 out of 15 counts in such a narrow energy range is highly suggestive of the presence of an emission line.  
The EPIC-pn spectrum can indeed be fitted by a narrow emission line at 0.78 keV also absorbed by a $N_\mathrm{H}$ of 5.4$\times10^{18}$ cm$^{-2}$.  
The X-ray flux in the 0.3 to 7.0 keV band would also be consistent with the value reported by C2022.  
\\

\section{Discussion and Concluding Remarks}

C2022 had to devote a major effort in their analysis of the Chandra observations of G\,29-38 to demonstrate that the data indeed implied a real detection of X-ray emission and that it could solely be attributed to this WD rather than to a cosmic background source.  
Our analysis of the XMM-Newton observation of G\,29-38 confirms the detection of a source at the 2005.91 proper-motion-corrected location of the WD with similar spectral properties and X-ray emission level as the source detected by Chandra at the 2020.73 proper-motion-corrected location of G\,29-38. 
This result provides strong support to the conclusions presented by C2022 confirming without any doubt the association of an X-ray source with G\,29-38.

The spectral shape of the XMM-Newton EPIC-pn spectrum is also consistent with that of the Chandra ACIS-S one.  
C2022 favored a plasma emission model with the photospheric chemical abundances of G\,29-38 with notable O and Fe enhanced abundances  \citep{Farihi_etal2009,Xu_etal2014}, which are consistent with those of its debris disk \citep{Reach_etal2009}. 
The spectral shapes of both X-ray observations are actually very highly indicative of line emission in the 0.7--0.8 keV range, which can be attributed to the O~{\sc viii} 16 \AA\ line or to the Fe complex at $\approx$16 \AA\ including emission lines of high excitation species from Fe~{\sc xvi} to Fe~{\sc xix}.  
The presence of these species would imply plasma temperature in the range from $\approx$2 to $\approx$8 MK, i.e., $\approx$0.17--0.7 keV.
At higher temperatures, the ionic fractional abundances of Fe shift towards higher ionization species whose emission lines peak at energies above 1 keV.  
Tests with absorbed, optically-thin thermal plasma {\it vapec} model with chemical abundances of the photosphere of G\,29-38 described by C2022 varying the plasma temperature indicated that the lowest possible plasma temperature is unconstrained, whereas the plasma temperature is certainly lower than 0.6 keV, in accordance with C2022 findings.

\begin{figure}
\begin{center}
\includegraphics[angle=0,width=\linewidth]{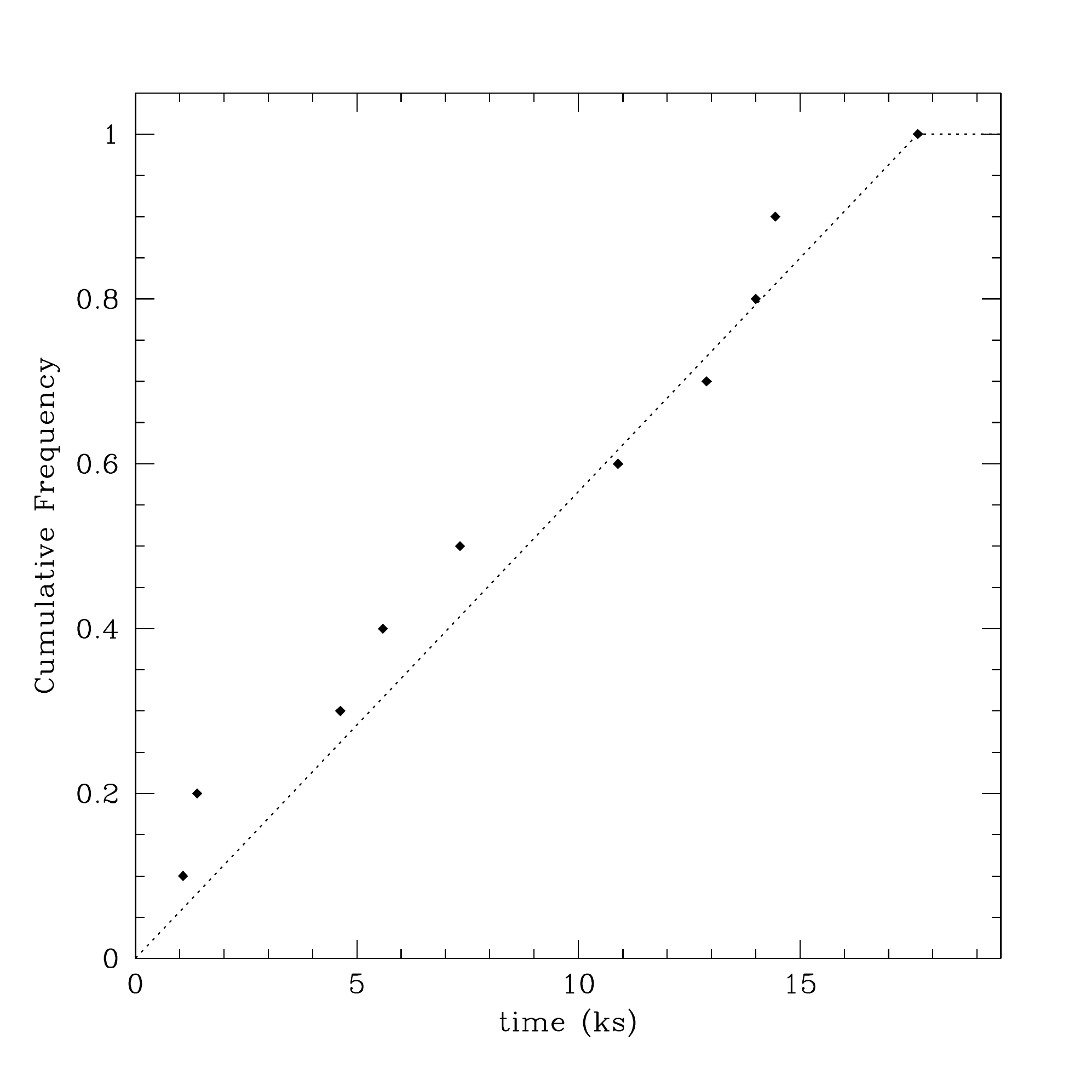}
\caption{Cumulative frequency of the X-ray photons in the 0.4--1.2 keV energy range detected in the aperture of G\,29-38. The dashed-line represents a constant X-ray flux.}
\label{fig:lc}
\end{center}
\end{figure}

The XMM-Newton EPIC-pn X-ray flux determined from these different spectral analyses also confirms that the level of X-ray emission of the source is similar, within their $\approx$40\% uncertainty, to that of the Chandra ACIS-S spectrum acquired almost 15 years apart.  
Although long-term variability is still possible, this consistency excludes the detection of a sudden accretion event either by the Chandra or the XMM-Newton observations.  
An inspection of the time arrival of the (few) photons detected by the XMM-Newton observations within the aperture of G\,29-38 in the range from 0.4 to 1.2 keV, which minimizes the contributions of the softer background and harder nearby source, finds that the measured count rate is consistent with a constant X-ray flux.

The possible variability of the accretion rate onto chemically polluted WDs, maybe involving stochastic discrete events of accretion, is an appealing idea \citep{Wyatt_etal2014,KB2017}, which can be also connected with the IR variability of dusty disks around them as they are replenished and subsequently depleted of material \citep{Swan_etal2019}. 
Indeed \citet{vHT2007} and then \citet{vHT_etal2009} reported variations in the photospheric Ca and Mg line-strengths of G\,29-38 attributed to episodic accretion events, but this result has been disputed \citep{DL-M2008}. 
Since the X-ray flux is mostly dependent on the accretion rate onto the WD, the steady count rate from G\,29-38 is consistent with a stable accretion rate.
Furthermore it argues against the presence of an accretion hot spot on the surface of G\,29-38, that would indicate channeling of infall material by a magnetic field, as has also been rejected in the case of GD\,394 \citep{Wilson_etal2019}.

The analysis of the archival XMM-Newton observations presented here confirms the X-ray emission from G\,29-38 and is consistent with a stable accretion rate on time scales of years and hours.  
The spectral shape is highly indicative of Fe and/or O rich material that would originate from a rocky planet debris. The high Fe abundances of the X-ray-emitting material is in line with the abundances of the dusty disk around G\,29-38 \citep{Farihi_etal2009}, which are then diminished in the stellar atmosphere as it settles in time-scales of a few weeks \citep{Xu_etal2014}.

The low luminosity $\leq 10^{26}$ erg~s$^{-1}$ (and thus accretion rate) and plasma temperature $\approx$0.17-0.7 keV of the X-ray emission from G\,29-38 is far from systems with high accretion rates such as symbiotic stars and CVs \citep[$L_{\rm X} \geq 10^{31}$ erg~s$^{-1}$ and $T_{\rm X} \geq$ 1 keV, see Figure~3 in][]{GCT2019}.  
The X-ray luminosity is also below that of putative single WDs with hard X-ray emission, which may still present plasmas at similar temperatures \citep[$L_{\rm X} \sim 5\times10^{29}-5\times10^{31}$ erg~s$^{-1}$ and $T_{\rm X} \sim$ 0.1--1.5 keV,][Estrada-Dorado et al., submitted to ApJ]{Chu2021}. 
The hard X-ray emission from these WDs, which is found to be variable \citep[e.g., with a period of 4.7 hr for KPD\,0005$+$5106,][]{Chu2021}, would arise from the accretion of material from a late-type stellar or a substellar companion. 
The low X-ray luminosity and plasma temperature, and the steady level of X-ray emission favor a bombardment solution \citep{KP1982}.   
\\
\\
\\

\noindent The authors are thankful to the anonymous referee for comments and suggestions that improved the presentation and interpretation of our results. S.E.-D. thanks Consejo Nacional de Ciencia y Tecnolog\'{i}a (CONACyT, Mexico) for a student scholarship. 
S.E.-D. and J.A.T. acknowledges funding by the Direcci\'on General de Asuntos del Personal Acad\'emico (DGAPA) of the Universidad Nacional Aut\'onoma de M\'exico (UNAM) project IA101622. J.A.T. thanks the Sistema Nacional de Investigadores (SNI-CONACyT, Mexico) and the Visiting-Incoming programme of the
IAA-CSIC through the Centro de Excelencia Severo Ochoa (Spain). 
M.A.G. and C.R.-L. acknowledge financial support from State Agency for Research of the Spanish MCIU through the ``Center of Excellence Severo Ochoa'' award to the Instituto de Astrof\'isica de Andaluc\'ia (SEV-2017-0709). M.A.G. thanks support from the Spanish Ministerio de Ciencia, Innovaci\'on y Universidades (MCIU) grant PGC2018-102184-B-100. Y.H.-C. acknowledges the grant MOST 110-2112-M-001-020 from the Ministry of Science and Technology (Taiwan).
This work has made extensive use of NASA's Astrophysics Data System.

%

\vspace{5mm}
\facilities{XMM-Newton\,(EPIC)}


\software{SAS\,\citep{Gabriel2004}, ESAS \citep{Snowden2004,Snowden2008,Kuntz2008}, XSPEC \citep{Arnaud1996}}




\end{document}